\documentclass[letterpaper,12pt,oneside,onecolumn,final,openright]{article}

\usepackage[utf8]{inputenc} 
\usepackage[T1]{fontenc}    
\usepackage[spanish]{babel} 

\usepackage{multicol}
\usepackage{enumerate}
\usepackage{amsmath, amsthm, amsfonts} 
\numberwithin{equation}{section}

\usepackage{amssymb}		

\usepackage{amsfonts}
\usepackage{latexsym}
\usepackage{anysize}
\marginsize{3.5cm}{2.5cm}{2.5cm}{2.5cm}
\usepackage[pdftex]{graphicx}
\usepackage{subfigure}
\usepackage{cite}
\usepackage{authblk}
\usepackage{longtable}
\usepackage{pdflscape}
\usepackage{graphicx}  
\begin{document}

\title{Sobre la naturaleza, tensorial o no tensorial, de los símbolos de Christoffel}

\author{Leonardo Patiño\footnote{leopj@ciencias.unam.mx}, Jessica Söhle\footnote{urkek.kanatlarinda@gmail.com}}
\affil{ Facultad de Ciencias, Universidad Nacional
Autónoma de México, A.P. 50-542, México D.F. 04510, México}

\maketitle

\begin{abstract}
Al estudiar un primer curso de Relatividad General y aprender sobre los símbolos de Christoffel, es fácil notar que sus componentes coordenadas no transforman como lo haría un tensor. Es por ello que resulta confuso encontrar libros de texto de alto prestigio en el área que se refieren a estos símbolos como si fueran un tensor, siendo un ejemplo notable de estos libros el escrito por Robert Wald \cite{GeneralRelativity}, que ha sido usado como libro de texto estándar en muchas instituciones a nivel mundial. La alternativa más simple sería suponer que los libros como los mencionados en el párrafo anterior describen a los símbolos de Christoffel de manera equivocada, pero esto no es el caso. Entender el fundamento matemático por el cual ciertos autores deliberadamente se refieren a los símbolos de Christoffel como si fueran un tensor aporta un entendimiento relevante de la naturaleza de la conexión métrica en Relatividad General, y es por ello que en este artículo nos hemos dado a la labor de presentar el análisis necesario. La discusión aquí presentada está enfocada a estudiantes que hayan tomado al menos un curso de relatividad general, así que algunas de las herramientas teóricas se considerarán conocidas.

\begin{center}
\textbf{Abstract}
\end{center}

From the first approach to General Relativity we learn about the Christoffel symbols, and it is easy to notice that their coordinated components do not transform as a tensor. It is then confusing to find renowned books on the subject that refer to these symbols as if they were indeed a tensor. A prominent example of such a book is the one written by Rober Wald \cite{GeneralRelativity}, which has been embraced as a textbook on the matter by a large amount of prestigious institutions around the globe. The simplest alternative would be to think that the books just mentioned present a wrong description of the Christoffel symbols, but this is not the case. Understanding the mathematical foundation of why some authors treat the Christoffel symbols as a tensor provides valuable insight about the nature of the metric connection in General Relativity, and that is why in this paper we dedicate the space to the needed analysis. The discussion provided here is aimed to those that have already taken at least one class in General Relativity, hence some of the theoretical tools will be taken as known by the reader.
\end{abstract}

Palabras Clave: Relatividad General, Símbolos de Christoffel, Tensores.

Keywords: General Relativity, Christoffel Symbols, Tensors

PACS: 04.20.-q, 02.40.Hw

\section{Introducción}

La esencia de la Relatividad General de Einstein radica en tratar a los fenómenos gravitacionales como una consecuencia de la geometría del espaciotiempo\cite{Misner:1974qy}. En esta teoría, que consideramos la mejor descripción de la gravitación a la fecha, la materia y energía provocan que el espaciotiempo se curve y a su vez, la materia en movimiento está obligada a seguir las trayectorias geodésicas del espaciotiempo mismo. Una geodésica se puede describir como una curva cuya tangente se transporta paralelamente a lo largo de si misma\cite{Semi-RiemannianGeometry}, pero para que el contenido de la frase anterior sea claro, es necesario entender en que consiste el transporte paralelo de vectores en un espaciotiempo curvo.

Transportar un vector conlleva a comparar vectores en puntos distintos del espaciotiempo, tarea que parecería ser sencilla si el espaciotiempo es plano, pero si elespaciotiempo es curvo, esta comparación requiere de alguna prescripción sobre como se modifican los vectores conforme nos desplazamos. Esta prescripción esta codificada en el objeto que llamamos conexión\cite{Baez:1995sj}. En particular, la conexión que es relevante en Relatividad General es la conexión métrica, que garantiza que el valor del producto interno, dictado por la métrica del espaciotiempo, entre dos vectores cualesquiera, no se modifique al trasportarlos usando esta conexión\cite{Misner:1974qy}.

Como vemos de los párrafos anteriores, la importancia de la conexión en el contexto de la Relatividad General es muy grande. Así mismo, un principio fundamental de la teoría es que su formulación sea independiente de las coordenadas que arbitrariamente elijamos para describirla, por lo que es muy importante conocer la forma en que los distintos objetos de relevancia para la teoría transforman ante un cambio de este tipo.

En este artículo estudiamos las propiedades que tienen los símbolos de Christoffel ante transformaciones de coordenadas a través de estudiar si son de carácter tensorial o no. El lector se dará cuenta a lo largo del presente escrito de que el análisis que se propone en la frase anterior es más sutil de lo que se pensaría a partir de una simple inspección de la expresión coordenada de los símbolos de Christoffel en términos de la métrica.

\section{Tensores}\label{tensores}

Iniciemos nuestra discusión recordando al lector la definición de tensor, seguido por una pequeña serie de comentarios sobre ella para ganar familiaridad con este tipo de objetos matemáticos y en particular con la forma operacional en que serán utilizados en este trabajo. Para el beneficio del lector que así lo requiera, pero para no disminuir la continuidad del argumento, se han incluido un número de apéndices con recordatorios sobre algunos temas necesarios de álgebra lineal y geometría diferencial, dejando en el texto principal solamente los puntos clave para la exposición.

Un tensor $T$ de rango $(m,n)$ es un operador multilineal que actúa sobre $m$ co-vectores, o vectores duales, y $n$ vectores, resultando en un número real \cite{LovelockTensor}, es decir,
\begin{equation}
T:V^*\times V^*\times ... \times V^*\times V\times ... \times V \rightarrow R,
\end{equation}
donde $V$, que se repite $n$ veces, es un espacio vectorial y $V^*$, que se repite $m$ veces, es el espacio dual al $V$.

El que este mapeo sea multilineal quiere decir que cumple con las siguiente propiedades. Si cualquiera de los vectores o co-vectores sobre los que actúa se escribe como una suma de dos vectores o co-vectores, el resultado de la acción de $T$ sobre este nuevo conjunto será la suma de las acciones sobre los conjuntos que tengan a cada vector o co-vector entre ellos, es decir, que la igualdad
\begin{eqnarray}
T(u^1,...,u^m,v_1,...,v_{i1}+v_{i2},...,v_n)&=&T(u^1,...,u^m,v_1,...,v_{i1},...,v_n)\nonumber \\
&+&T(u^1,...,u^m,v_1,...,v_{i2},...,v_n),\label{ml1}
\end{eqnarray}
se cumple independientemente de la entrada en que se reemplace el elemento por una suma. La otra propiedad a cumplir es que si cualquiera de los vectores o co-vectores sobre los que actúa se reemplaza por un múltiplo de si mismo, el resultado de la acción de $T$ sobre este nuevo conjunto será el mismo múltiplo de veces el resultado de actuar con $T$ sobre el conjunto original de vectores y co-vectores, es decir, que la igualdad
\begin{equation}
T(u^1,...,u^m,v_1,...,\alpha v_i,...,v_n)=\alpha T(u^1,...,u^m,v_1,...,v_i,...,v_n),\label{ml2}
\end{equation}
se cumple independientemente de la entrada en que se multiplique por la constante $\alpha$.

La noción de que un tensor es un objeto parecido a una matriz, pero probablemente con más índices, tiene que ver con que una matriz $M$ puede tomar un vector renglón $V_1$ y un vector columna ${V_2}^T$, que en este caso toma el lugar de un co-vector, y mandarlos multilinealmente a un real por medio de la operación matricial $V_1 M {V_2}^T$. Es en este sentido que una matriz puede actuar como un tensor de rango (1,1).

De particular importancia para la discusión central de este artículo es que de la definición de tensor vista arriba se sigue que un mapeo lineal $\cal{T}$ de $V$ al espacio $\mathbf{T}^{(m,n)}$ de tensores de rango $(m,n)$,
\begin{equation}
{\cal{T}}:V\rightarrow\mathbf{T}^{(m,n)},
\end{equation}
es un tensor de rango $(n,m+1)$.

Para ver que la aseveración del párrafo anterior es cierta, basta con notar que tras haber actuado linealmente sobre un elemento de $V$, es decir, sobre un vector $v$, ${\cal{T}}(v)$ es un mapeo multilineal capaz de actuar sobre $m$ co-vectores y $n$ vectores. De lo anterior sigue que $\cal{T}$ es en sí un mapeo que actúa multilinealmente sobre $m$ co-vectores y $n+1$ vectores en total para arrojar un número real, satisfaciendo en consecuencia la definición misma de un tensor de rango $(n,m+1)$.

De la misma forma, un mapeo multilineal
\begin{equation}
{\cal{T}}:V^*\times V^*\times ... \times V^*\times V\times\times ... \times V \rightarrow \mathbf{T}^{(m,n)},
\end{equation}
donde $V^*$ se repite $M$ veces y $V$ se repite $N$, es un tensor de rango $(M+m,N+n)$.

Respecto a la forma de escribir a un tensor, recordando que un vector $v$ de un espacio $V$ provee un mapeo lineal del espacio dual $V^*$ a los reales\footnote{Para recordar como un vector puede actuar sobre un co-vector se puede consultar el apéndice \ref{dual}}, nos damos cuenta de que el espacio $\mathbf{T}^{(m,n)}$ de tensores de rango $(m,n)$ es el producto tensorial de $m$ veces $V$ y $n$ veces $V^*$, de tal forma que un tensor de rango $(m,n)$ se puede escribir usando la base de vectores $e_i$ y su base dual $e^j$ como
\begin{equation}
T={T^{i_1i_2...i_m}}_{j_1j_2...j_n}e_{i_1}\times e_{i_2}\times...\times e_{i_m}\times e^{j_1}\times e^{j_2}\times...\times e^{j_n},\label{tenscomp}
\end{equation}
donde cada uno de los multiíndices $i_a,j_b$ toman valores entre $1$ y $D$, con $D$ la dimensionalidad de $V$, que es la misma que la de $V^*$. La forma de entender (\ref{tenscomp}) es notando que cada uno de los vectores $e_{i_a}$ actúa linealmente sobre un co-vector arbitrario y resulta en un real, así como cada uno de los co-vectores $e^{j_b}$ actúa linealmente sobre un vector arbitrario y resulta en un real. Notamos así que cada uno de los términos $e_{i_1}\times e_{i_2}\times...\times e_{i_m}\times e^{j_1}\times e^{j_2}\times...\times e^{j_n}$ en (\ref{tenscomp}), es un objeto que actúa multilinelamente sobre $n$ co-vectores y $m$ vectores arbitrarios.

No es difícil ver que en la suma involucrada en (\ref{tenscomp}) hay $D^{m+n}$ elementos $e_{i_1}\times e_{i_2}\times...\times e_{i_m}\times e^{j_1}\times e^{j_2}\times...\times e^{j_n}$, y que ellos constituyen una base para el espacio de tensores $\mathbf{T}^{(m,n)}$, por lo que decimos entonces que los coeficientes ${T^{i_1i_2...i_m}}_{j_1j_2...j_n}$ son las componentes del tensor $T$ en esta base.

El resultado de la acción de un tensor de rango $(m,n)$ sobre $m$ co-vectores y $n$ vectores dados debe de ser independiente de la base $e_i$ (y el dual $e^j$) que se elija para representar a estos objetos. Esta observación implica que si hay dos bases distintas de vectores $e_i$ y $\tilde{e}_{\tilde{i}}$ que se escriben una respecto a la otra según $\tilde{e}_{\tilde{i}}={M^i}_{\tilde{i}}e_i$, las componentes de un tensor $T$ respecto a la base $e_i$ y respecto a la base $\tilde{e}_{\tilde{i}}$ deben inevitablemente estar relacionadas de acuerdo a
\begin{equation}
{\tilde{T}^{\tilde{i_1}\tilde{i_2}...\tilde{i_m}}}_{\tilde{j_1}\tilde{j_2}...\tilde{j_n}}={M^{j_1}}_{\tilde{j_1}}{M^{j_2}}_{\tilde{j_2}}\ldots {M^{j_n}}_{\tilde{j_n}}{(M^{-1})^{\tilde{i_1}}}_{i_1}{(M^{-1})^{\tilde{i_2}}}_{i_2}\ldots{(M^{-1})^{\tilde{i_m}}}_{i_m}{T^{i_1i_2...i_m}}_{j_1j_2...j_n},\label{trans}
\end{equation}
pues solo así el resultado de la acción de $T$ sobre los vectores y co-vectores escritos en términos de la base $e$ será igual al resultado de $\tilde{T}$ actuando sobre los vectores y co-vectores escritos en términos de la base $\tilde{e}$.

Queda claro entonces que un objeto cuyas componentes transformen de forma distinta a (\ref{trans}) no puede ser un tensor.

\section{Tensores en Relatividad General}

Hasta el momento hemos hablado sólo de vectores, co-vectores y tensores construidos sobre un espacio $V$ y su dual, pero, dado que estamos interesados en el formalismo de la relatividad general, debemos recordar que en ella los vectores existen en los espacios tangentes al espacio-tiempo en cada punto $p$ de este\cite{choquet}. Dicho de otra forma, cada punto $p$ tiene asociado un espacio vectorial $V$ dado por el espacio tangente en ese punto. Cada uno de estos espacios vectoriales se puede usar para construir el espacio dual $V^*$ y con ello, los espacios tensoriales de cualquier rango.

Llamemos entonces $v_p$ a los elementos del espacio tangente en el punto $p$, y $u_p$ a los elementos del espacio vectorial dual a este, llamado también espacio cotangente. El objeto que asocia un vector $v_p$ a cada punto $p$ le llamamos campo vectorial\footnote{Una definición operacional de campo vectorial se puede encontrar en el apéndice (\ref{Apcon})}, y por simplicidad lo denotaremos $v$. De igual forma podemos construir campos de vectores duales $u$ que asocian un $u_p$ a cada punto $p$, y a estos campos los llamaremos 1-formas. Continuando en esta dirección, un campo tensorial debe de asociar un tensor a cada punto del espacio-tiempo, pero un tensor esta definido por el mapeo multilineal discutido en la sección anterior, así que un campo tensorial $T$ de rango $(n,m)$ debe ser un objeto que actúa multilinealmente sobre $n$ 1-formas y $m$ campos vectoriales.

El resultado de actuar con una 1-forma $u$ sobre un campo vectorial $v$ es fácil de visualizar, pues el co-vector asociado a cada punto $u_p$ debe actuar sobre el vector asociado al mismo punto $v_p$ y arrojar un número real. Un objeto que asocia un número real a cada punto del espacio es una función real, así que una 1-forma actúa sobre un campo vectorial y resulta en una función real. Para respetar la linealidad de la acción en cada punto que debe cumplir cualquier co-vector al operar sobre un vector, es necesario que las relaciones 
\begin{eqnarray}
u(v_1+v_2)&=&u(v_1)+u(v_2)\,\,\,\,\,\,\,\,\,\,\,\,\mathrm{y} \label{lin1f1}\\
u(f\, v)&=&fu(v)\label{lin1f2}
\end{eqnarray}
se satisfagan para cualesquiera vectores $v,v_1,v_2$ y función $f$ arbitrarios.

De la misma forma, la acción de un campo tensorial $T$ sobre las 1-formas y los campos vectoriales correspondiente a su rango debe ser una función real. Para implementar la multilinealidad de $T$ en este contexto, notamos que (\ref{ml1}) y (\ref{ml2}) se deben cumplir en todo punto $p$ del espacio-tiempo, y considerando que una función real $f$ asocia un número real $\alpha$ a cada $p$, no es difícil convencerse de que las relaciones de multilinealidad que se deben cumplir para un campo tensorial son una idéntica a (\ref{ml1}) con 1-formas y campos vectoriales y otra dada por
\begin{equation}
T(u^1,...,u^m,v_1,...,f v_i,...,v_n)=f T(u^1,...,u^m,v_1,...,v_i,...,v_n).\label{mlcamp}
\end{equation}

Vemos entonces que (\ref{mlcamp}) se reduce a (\ref{ml2}) al ser evaluada en un punto particular $p$, pues $f(p)=\alpha$, pero aún más importante es que $\alpha$ está determinado exclusivamente por el valor de $f$ en el punto $p$, y para este valor es irrelevante el comportamiento de $f$ en una vecindad de $p$. Nos referimos a este comportamiento como que $T$ es multilineal sobre el espacio de las funciones.

\section{Conexiones}\label{conexiones}

Una conexión $D$ para el haz tangente al espacio-tiempo es un operador\cite{Baez:1995sj} que a cada campo vectorial $v$ le asocia un mapeo del espacio de los campos vectoriales, que denotaremos como $\cal{V}$, en si mismo,
\begin{equation}
D_v:\cal{V}\rightarrow \cal{V},\label{conn}
\end{equation}
que cumple con las siguientes propiedades
\begin{eqnarray}
D_v(\alpha v_1)&=&\alpha D_v(v_1),\\
D_v(v_1+v_2)&=&D_v(v_1)+D_v(v_2),\label{lienalsuma}\\
D_v(f v_1)&=&v(f)v_1+f D_v(v_1),\label{leibniz}\\
D_{v+u}(v_1)&=&D_v(v_1)+D_u(v_1),\label{linealv1} \\
D_{f v}(\alpha v_1)&=&f D_v(v_1), \label{linealv2}
\end{eqnarray}
con $v,u,v_1$ y $v_2$ campos vectoriales arbitrarios, $\alpha$ cualquier constante y $f$ una función arbitraria también\footnote{En el apéndice \ref{Apcon} se recuerda que $v(f)$ es la derivada de la función $f$ en la dirección de $v$}.

Vemos entonces que una conexión es un objeto que tras actuar sobre dos campos vectoriales, $v$ y $v_1$ en las ecuaciones anteriores, resulta en otro campo vectorial dado por $D_v(v_1)$. El que $D_v(v_1)$ sea un campo vectorial implica que es un objeto que está listo para actuar linealmente sobre cualquier 1-forma y arrojar una función cumpliendo con las igualdades,
\begin{eqnarray}
D_v(v_1)[\omega_1+\omega_2]&=& D_v(v_1)[\omega_1]+D_v(v_1)[\omega_2],\,\, \mathrm{y}\label{linealw1}\\
D_v(v_1)[f \omega]&=&f D_v(v_1)[\omega],\label{linealw2}
\end{eqnarray}
para una función arbitraria $f$.

Del párrafo anterior vemos que es factible visualizar a una conexión como un objeto que es capaz de actuar sobre dos campos vectoriales y una 1-forma resultando en una función. A pesar de esta observación, esto no implica que una conexión sea un tensor, dado que su acción no es lineal para todos los objetos sobre los que actúa. En particular, aunque (\ref{linealv1}), (\ref{linealv2}), (\ref{linealw1}) y (\ref{linealw2}) indican que la acción de la conexión es lineal sobre el espacio de las funciones para dos de los objetos sobre los que actúa, la propiedad (\ref{leibniz}) muestra que la acción de la conexión no cumple con ser lineal para el tercer objeto.

Es fácil ver como una conexión sobre el haz tangente determina a una conexión tanto en el haz cotangente como sobre los espacios de campos tensoriales de rangos arbitrarios. Esta generalización no es necesaria para el desarrollo del argumento de este artículo, pero para aquellos lectores que deseen un breve recordatorio de estos detalles, así como de la definición operacional de un campo vectorial, hemos incluido el apéndice \ref{Apcon}. En general en lo que resta del artículo cuando hablemos de una conexión, nos referiremos a una conexión para el haz tangente al espacio-tiempo.

De importancia central al argumento de este artículo es recordar\cite{Baez:1995sj} que dada una conexión $D$, cualquier otra conexión $D'$ se puede escribir sumándole a la primera una 1-forma $A$ cuyos componentes son mapeos lineales del espacio tangente en si mismo,
\begin{equation}
D'=D+A.\label{A}
\end{equation}
Para que (\ref{A}) tenga sentido, recordemos que las 1-formas con entradas reales, que son las que hemos visto hasta ahora, al actuar sobre un campo vectorial, resultan en una función, es decir, un objeto que asocia un real a cada punto del espacio base. De forma análoga al párrafo anterior, una 1-forma cuyos componentes son mapeos lineales es un objeto que al actuar sobre un campo vectorial $v$ da lugar a $A_v$, que es un objeto que asocia un mapeo lineal para cada punto de la variedad base. La información que $A$ codifica es la forma en que se deben transportar los vectores del espacio tangente, representada por la acción de la transformación lineal $A_v$ sobre de ellos, al desplazarse en la dirección dada por $v$, para acomodar la diferencia entre las conexiones. En breve confirmaremos que en efecto $A$ es una 1-forma cuyos componentes son campos de mapeos lineales.

Más adelante nos daremos cuenta de que la pregunta relevante a la presente discusión es respecto a la naturaleza de $A$, en particular si es un tensor o no, así que analicemos con cuidado los elementos que la constituyen.

Un mapeo lineal\cite{Bretscher} del espacio tangente en si mismo, es un objeto que actúa linealmente sobre un vector y resulta en otro vector. Un objeto $L$ que asocia un mapeo lineal a cada punto del espacio tiempo es aquel que actúa linealmente sobre un campo vectorial y resulta en otro campo vectorial, $L(v)=u$, donde a cada campo vectorial $v$ le corresponde un campo vectorial $u$ específico. Las condiciones de linealidad son las usuales
\begin{eqnarray}
L(v_1+v_2)&=&L(v_1)+L(v_2)\,\,\,\,\,\,\,\,\, \mathrm{y} \nonumber \\
L(f\, v)&=&f\, L(v) \label{maplin1}
\end{eqnarray}
con $f$ una función arbitraria.

Como habíamos mencionado antes, un vector es un mapeo lineal del espacio de los covectores a los reales. El objeto resultante de actuar con un mapeo lineal sobre un vector es otro vector, y por lo tanto este objeto resultante es un mapeo lineal de los covectores a los reales. De lo anterior concluimos que un objeto $L$ que asocia un mapeo lineal a cada punto del espacio tiempo, tras actuar sobre un campo vectorial $v$, es otro campo vectorial $L(v)=u\in {\cal{V}}$, y por lo tanto $L(v)$ es un objeto listo para actuar linelamente sobre una 1-forma $\omega$ y arrojar una función $f$,  $$L(v)[\omega]=f,$$ donde a cada co-vector $\omega$ le corresponde una función específica $f$.

A partir de (\ref{maplin1}) y dada la linealidad de la acción de $L(v)$ sobre $\omega$, vemos que podemos pensar a $L$ como un mapeo multilineal que actúa sobre un campo vectorial $v$ y una 1-forma $\omega$, y resulta en una función, es decir, podemos pensar en $L$ como un campo tensorial de rango $(1,1)$.

Hemos dicho, sin aún demostrarlo, que $A$ es una 1-forma cuyos componentes son mapeos lineales, pero si este es el caso, vemos que $A$ es un tensor de rango $(1,2)$, pues $A$, tras haber actuado linealmente sobre un campo vectorial $v$, resulta en $A_v$, un mapeo lineal en cada punto del espacio, que según lo visto en el párrafo anterior, es un campo tensorial de rango $(1,1)$, con lo que en total $A$ actúa multilinealmente sobre dos campos vectoriales y una 1-forma, resultando en una función escalar.

Para finalmente ver que $A$ es una 1-forma cuyos componentes son mapeos lineales, notemos que (\ref{A}) puede ser reescrito como $A=D'-D$. A partir de las propiedades (\ref{conn}) a (\ref{linealv2}) vemos que si bien el resultado de actuar con $D$ sobre los campos vectoriales $v_i$ es otro vector, también notamos, en particular a partir de (\ref{leibniz}), que este no es un mapeo lineal de $\cal{V}$ en si mismo. Para demostrar que sin embargo la acción de $A$ sobre los vectores $v_i$ sí es la de una transformación lineal en cada punto del espacio basta con verificar dos propiedades, una es que $A_v(v_i)$ es lineal sobre el espacio de las funciones en su acción sobre $v_i$, y la otra es que el resultado de $A_v(v_i)$ en el punto $p$ sólo depende del vector tangente $({v_i})_p$, y no del comportamiento del campo vectorial $v_i$ en la vecindad de $p$.

La propiedad de linealidad de la acción de $A_v$ sobre la suma de vectores está garantizada por (\ref{lienalsuma}), y para verificar la linealidad sobre el espacio de las funciones basta ver que
\begin{eqnarray}
A_v(f\, v_i)&=& \left[ {D'}_v-D_v\right] (f\, v_i)\nonumber \\
&=& {D'}_v  (f\, v_i) - D_v (f\, v_i)\nonumber  \\
&=& (\partial_v (f)\, v_i + f \, {D'}_v(v_i)) -(\partial_v (f)\, v_i + f\, D_v (v_i))\label{linA} \\
&=& f \, {D'}_v(v_i) - f\, D_v (v_i)\nonumber \\
&=& f \left[ {D'}_v-D_v\right] (v_i)\nonumber \\
&=& f A_v(v_i).\nonumber 
\end{eqnarray}

La segunda propiedad que queremos comprobar es que $[A_v(v_1)]_p=[A_v(v_2)]_p$ para cualesquiera dos campos vectoriales $v_1$ y $v_2$ que cumplan con ser iguales en el punto $p$, pero en general diferentes en una vecindad alrededor de este. Primero reescribimos  $[A_v(v_1)]_p=[A_v(v_2)]_p$ como  $[A_v(v_1)]_p-[A_v(v_2)]_p=0$ y notamos que dadas (\ref{conn}) a (\ref{linealv2}) en general
\begin{eqnarray}
A_v(v_1)-A_v(v_2)&=&\left[ {D'}_v-D_v\right] (v_1)-\left[ {D'}_v-D_v\right] (v_2)\\
&=&\left[ {D'}_v-D_v\right] (v_1-v_2).
\end{eqnarray}
Dado que la suma o la resta de dos campos vectoriales es un campo vectorial, $v_1-v_2$ cae en esta categoría y puede ser escrito como $v_1-v_2=g\, u$, para algún campo no nulo $u$ y una función $g$ que debe cumplir con $g(p)=0$, dado que $({v_1})_p=({v_2})_p$ y dado que $u(p)\neq 0$, por ser no nulo. Continuamos con el cálculo sustituyendo $g\, u$ en lugar de $v_1-v_2$ y echando mano de (\ref{linA}),
\begin{eqnarray}
\left[ {D'}_v-D_v\right] (v_1-v_2)&=&\left[ {D'}_v-D_v\right] (g\, u)\nonumber \\
&=& g \left[ {D'}_v-D_v\right] (u),\nonumber
\end{eqnarray}
y concluimos así que
\begin{equation}
A_v(v_1)-A_v(v_2)=g \left[ {D'}_v-D_v\right] (u).\label{local}
\end{equation}
Puesto que $g(p)=0$, la evaluación de (\ref{local}) en $p$ demuestra que $[A_v(v_1)]_p=[A_v(v_2)]_p$ siempre que $v_1$ y $v_2$ sean iguales en $p$, independientemente de su comportamiento en la vecindad de el.

Con esto hemos verificado que $({A_v})_p$ es un mapeo lineal, del espacio tangente al punto $p$ en si mismo, y dadas las propiedades de linealidad respecto a $v$ en (\ref{conn}) - (\ref{linealv2}), queda demostrado que $A$ en efecto es una 1-forma con mapeos lineales por componentes y con ello que $A$ asigna a cada punto $p$ un tensor de rango $(1,2)$. Dicho de otra forma, $A$ es un campo tensorial de rango $(1,2)$.

\section{La derivada covariante}

Con lo que se ha establecido en las secciones anteriores nos encontramos en posición de exponer los resultados principales de este trabajo, sin embargo, para que estos se aprecien con la importancia que tienen, es necesario explicar el contexto físico en que son relevantes.

Uno de los dos paradigmas principales de la Relatividad General es notar que lo que percibimos como fuerza gravitacional es en realidad un efecto de la curvatura del espaciotiempo. Las implicaciones de que el espacio en que se desarrolla la física sea curvo son muy amplias, pero nos enfocaremos ahora en los de mayor relevancia para la presente discusión.

Los vectores tangentes en un punto $p_1$ del espaciotiempo, que representan cantidades físicas de mucha importancia, existen en el espacio tangente a $p_1$. Cuando el espaciotiempo es curvo, no hay una forma obvia de identificar a los elementos del espacio tangente en un punto $p_1$ con los del espacio tangente en otro punto $p_2$, pues, aunque isomórficos, no son el mismo espacio. En el caso plano la observación anterior es cierta también, pero la planitud del espacio hace posible imaginar una manera de identificar los vectores tangentes de puntos diferentes. En términos pictóricos, la noción de transportar un vector de $p_1$ a $p_2$ sin modificar ni la dirección en la que apunta ni su magnitud parece intuitiva e incluso podemos pensar en codificarla como que la derivada del vector en la dirección en que se está trasladando sea cero.

Nos gustaría extender esta idea intuitiva de trasportar un vector sin modificarlo al caso de los espacios curvos, así que veamos como la conexión introducida en la sección anterior nos ayuda a ello.

Lo primero por notar es que la conexión $D_v(v_1)$ actúa sobre $v_1$ como un operador de derivación, lo cual es particularmente claro a partir de (\ref{leibniz}). Parecería entonces que demandar $D_{\gamma'}(v_1)=0$ a lo largo de una curva $\gamma$ que conecta $p_1$ con $p_2$ y cuyo vector tangente es $\gamma'$ en cada punto de la trayectoria, garantizaría el que el vector $v_1$ se transporte sin cambio al recorrer $\gamma$. A este proceso se le llama transporte paralelo del vector $v_1$ a lo largo de $\gamma$.

Como vimos en la sección anterior, la conexión dista mucho de ser única, pues tenemos tanta libertad para elegirla como la que hay para fijar a $A$. Parecería entonces que no hemos logrado mucho, pues lo que entendemos por trasportar paralelamente a un vector es tan arbitrario como la elección de $A$, así que la única forma en que una conexión nos puede ayudar a determinar una manera de comparar vectores en puntos distintos es si existe un criterio que seleccione una $A$ particular como la mejor elección.

En Relatividad General existe un criterio para preferir a una $A$ sobre las demás, pues esta es una teoría geométrica, lo que atribuye un carácter de primordial importancia a la cantidad  $g(v_1,v_2)$. Si hemos de considerar que dos vectores no cambian al transportarse paralelamente, una cantidad asociada a ellos tan fundamental como lo es $g(v_1,v_2)$ no debe de cambiar al darse este tipo de transporte para $v_1$ y $v_2$ a lo largo de una trayectoria arbitraria. Lo interesante de esta observación es que satisfacer el criterio recién descrito es suficiente para determinar de forma única a la conexión que se debe usar.

El criterio del párrafo anterior se reduce a encontrar aquella $A$ tal que baste que se cumpla $D_{\gamma'}(v_1)=D_{\gamma'}(v_2)=0$ para garantizar que $D_{\gamma'}g(v_1,v_2)=0$. La forma más rápida de demostrar que esta $A$ es única es probablemente introduciendo índices coordenados, en términos de los cuales $D_{\gamma'}g(v_1,v_2)$ se escribe como $$\gamma'^\mu D_\mu(g_{\nu\sigma}v_1^\nu v_2^\sigma),$$ y dado que $D$ actúa sobre un producto de acuerdo a la ley de Leibnitz, las condiciones $D_{\gamma'}(v_1)=D_{\gamma'}(v_2)=0$ implican $$\gamma'^\mu v_1^\nu v_2^\sigma D_\mu(g_{\nu\sigma})=0.$$ Esta última condición se cumple para cualesquiera vectores transportados paralelamente a lo largo de cualquier curva si y sólo si \begin{equation}
D_\mu(g_{\nu\sigma})=0.\label{cova}
\end{equation}
Lo que queda por hacer es mostrar que (\ref{cova}) define de forma única a $D$ y en consecuencia a $A$ también, claro que, dado que los objetos así definidos son especiales, les daremos un nombre distinto, que será $\nabla$ a la primera y $\Gamma$ a la segunda para el caso en que escribamos la acción de $\nabla$ en comparación con la derivada parcial coordenada, es decir, $\nabla=\partial+\Gamma$.

Considerando que, como vimos en la sección anterior, $\Gamma$ es una 1-forma con valores endomórficos, podemos escribirla como la combinación lineal $\Gamma=(\Gamma^\alpha_\beta)_\delta (e_\alpha\otimes e^\beta)\otimes e^\delta$ de los elementos de las bases coordenadas $e_\alpha$ y $e^\beta$ de vectores y 1-formas respectivamente. En la expresión anterior los paréntesis no son relevantes, pero los incluimos para hacer obvio que parte es la endomórfica y cual la de 1-forma. Usando estas componentes la acción de $\nabla$ sobre $g_{\nu\sigma}$ se puede escribir como
\begin{equation}
\nabla_\mu(g_{\nu\sigma})=\partial_\mu (g_{\nu\sigma})- (\Gamma^\alpha_\nu)_\mu g_{\alpha\sigma}-(\Gamma^\alpha_\sigma)_\mu g_{\alpha\nu},
\end{equation} 
que en vista de (\ref{cova}) implica
\begin{equation}
\partial_\mu (g_{\nu\sigma})= (\Gamma^\alpha_\nu)_\mu g_{\alpha\sigma}+(\Gamma^\alpha_\sigma)_\mu g_{\alpha\nu},\label{partgamma}
\end{equation}
y permutando índices también
\begin{eqnarray}
\partial_\nu (g_{\sigma\mu})&=& (\Gamma^\alpha_\sigma)_\nu g_{\alpha\mu}+(\Gamma^\alpha_\mu)_\nu g_{\alpha\sigma},\\
\partial_\sigma (g_{\mu\nu})&=& (\Gamma^\alpha_\mu)_\sigma g_{\alpha\nu}+(\Gamma^\alpha_\nu)_\sigma g_{\alpha\mu}.
\end{eqnarray}

Usando la combinación lineal adecuada de las ecuaciones anteriores podemos escribir
\begin{equation}
\Gamma^\mu_{\nu\sigma}=\frac{1}{2}g^{\mu\sigma}\left( \partial_\nu g_{\sigma\sigma}+\partial_\sigma g_{\sigma\nu}-\partial_\sigma g_{\nu\sigma}\right) , \label{chrismet}
\end{equation}
relación que define de manera única a todos los coeficientes $\Gamma^\mu_{\nu\sigma}$ y en consecuencia a $\Gamma$ misma, demostrando no sólo que es posible construirla de manera que cumpla con los requisitos discutidos previamente, sino que además, es única.

La ecuación (\ref{chrismet}) es la expresión que permite calcular a los $\Gamma^\mu_{\nu\sigma}$, conocidos como los símbolos de Christoffel, en términos de los componentes de la métrica. 

Antes de llegar a los resultados principales del presente artículo, queremos insistir en la importancia de la conexión covariante, pues el preservar a la métrica ante el trasporte paralelo es lo que hace posible que la determinación de distancias pueda ser consistente en toda la extensión del espaciotiempo. Aún más, toda conexión confiere una cierta curvatura al espacio sobre el que está definida, pero la conexión covariante define a la curvatura que resulta de preservar la estructura métrica, convirtiéndose así en la curvatura relevante para Relatividad General y el elemento que permite escribir la ecuación de Einstein. Es por ello que es tan importante entender a fondo la naturaleza de esta conexión y los símbolos de Christoffel que están inevitablemente atados a ella.

\section{Los símbolos de Christoffel}

Como vimos en la sección anterior, los símbolos de Christoffel $\Gamma^\mu_{\nu\sigma}$ son los componentes de $\Gamma$, que es un $A$ muy particular dado por $\nabla=\partial+\Gamma$, con $\partial$ la derivada coordenada y $\nabla$ la derivada que cumple con la muy especial propiedad de conservar a la métrica ante el transporte paralelo. De esta aseveración, el punto relevante para esta sección y para el objetivo central del presente trabajo, es que los símbolos de Christoffel son un caso particular de $A$ y por ello, según lo que demostramos en la sección (\ref{conexiones}), representan un campo tensorial de rango $(1,2)$.

Esto parecería estar en clara contradicción con la expresión (\ref{chrismet}) que encontramos para calcular los símbolos de Christoffel a partir de la métrica, pues esta no transforma como un tensor ante cambios de coordenadas y, de acuerdo a lo que revisamos en la sección (\ref{tensores}), esto implicaría que $\Gamma$ no es un tensor.

El que (\ref{chrismet}) no transforma como tensor es fácil de ver, incluso con un ejemplo simple, como es el caso de un espacio plano dos dimensional.

Primero, por cálculo directo notamos que si la métrica en coordenadas cartesianas,
\begin{equation}
g_{\mu\nu}=
\begin{pmatrix}
1&0\\
0&1
\end{pmatrix},
\end{equation}
 se usa en (\ref{chrismet}) obtenemos que todos los componentes de los símbolos de Christoffel se anulan, mientras que si usamos la métrica polar,
\begin{equation}
\tilde{g}_{\mu\nu}=
\begin{pmatrix}
1&0\\
0&r^2
\end{pmatrix},
\end{equation} 
para el mismo espacio y la insertamos en (\ref{chrismet}) tendremos que hay tres componentes de los símbolos de Christoffel,
\begin{equation}
{\Gamma^\theta}_{r\theta}=\frac{1}{r},\,\,\,\,\,{\Gamma^\theta}_{\theta r}=\frac{1}{r}\,\,\,\,\,\mathrm{y}\,\,\,\,\,{\Gamma^r}_{\theta\theta}=-r,\label{chrispol}
\end{equation}
que son distintos de cero.

Para contrastar con el cálculo directo, queremos partir de los símbolos de Christoffel obtenidos al usar la métrica cartesiana en (\ref{chrismet}) y aplicarles una transformación tensorial de coordenadas para comparar el resultado con (\ref{chrispol}).

La trasformación entre coordenadas cartesianas y polares $x=r\cos(\theta),\,\,y=r\,{\mathrm{sen}}(\theta),$ da lugar a la transformación entre las componentes en términos de las bases de vectores coordenados 
\begin{equation}
{M^i}_{\tilde{i}}=
\begin{pmatrix}
\cos(\theta) & -r\,{\mathrm{sen}}(\theta) \\
{\mathrm{sen}}(\theta) & r\cos(\theta)
\end{pmatrix},\label{M}
\end{equation}
la cual está bien definida en todo lugar salvo en el origen.

Usando (\ref{M}) y su inversa en (\ref{trans}) vemos que el aplicar la transformación a los símbolos de Christoffel nulos provenientes de la métrica cartesiana resulta en un nuevo conjunto de coeficientes que son nulos también, salvo en el origen donde no está definida la transformación, y por ello ahí no podemos comparar los coeficientes.

Con esto vemos que los resultados del cálculo directo de (\ref{chrismet}) en distintas coordenadas no están relacionados por una transformación tensorial. Como se mencionó antes, esto parece conducirnos a una contradicción irreconciliable, de cuyo origen hablamos a continuación.

\section{Sobre la naturaleza tensorial o falta de ella en los símbolos de Christoffel}

La solución a la tensión entre los resultados expuestos en lo anterior de este escrito se resuelve más fácilmente de lo que parecería a primera vista. El punto central es que no existe tal tensión, pues ambos resultados son correctos y a continuación veremos porque.

Para estudiar al objeto $\Gamma$, examinemos a los tres elementos involucrados en la relación que lo define
\begin{equation}
\nabla=\partial+\Gamma. \label{defgamma}
\end{equation}
Notemos primero que $\nabla$ es el operador derivada covariante, dado por la conexión de Levi-Civita, que es única para cada métrica y cuya acción específica sobre vectores es independiente de las coordenadas en las que se hagan los cálculos. En contraste, $\partial$ es un objeto que denota el cálculo del cambio infinitesimal de cualquier cantidad, en particular, de un vector, al ser evaluado en puntos desplazados infinitesimalmente a lo largo de las direcciones dictadas por el sistema coordenado. Es por esta característica de $\partial$ que, aunque sea común escribirlo como el mismo objeto en cualquiera que sea el sistema de coordenadas que se esté usando, en realidad $\partial$ denota un objeto matemático distinto en cada sistema coordenado.

De forma concreta, $\nabla$ es el mismo objeto en todo sistema de coordenadas y simplemente tiene distintas representaciones acorde a los distintos sistemas coordenados, mientras que $\partial$ es un objeto diferente para cada sistema coordenado. Al ser $\Gamma$ definido en términos de estos dos objetos, es claro que $\Gamma$ mismo no es un único objeto para todos los sistemas de coordenadas, sino que para cada sistema existe un objeto $\Gamma$ no necesariamente igual para todos ellos. 

La fórmula (\ref{chrismet}) proporciona los componentes para el objeto $\Gamma$ correspondiente al sistema coordenado que use a $g$ como métrica, que, reiterando, no tiene por que ser el mismo objeto que el $\Gamma$ de otros sistemas coordenados. Siendo los componentes obtenidos a partir de (\ref{chrismet}) en distintos sistemas coordenados los componentes de objetos simplemente diferentes, estos no están obligados a guardar relación alguna entre ellos.

De esta manera vemos que no existe una contradicción en los resultados de secciones previas, pues si bien $\Gamma$ es un tensor, el tensor que llamamos $\Gamma$ puede ser un tensor distinto en dos sistemas de coordenadas diferentes, hecho que se refleja en que los componentes dados por (\ref{chrismet}) en diversos sistemas coordenados, siendo los componentes de tensores distintos, no se relacionan por medio de una transformación tensorial.

La discusión anterior concluye nuestros argumento, con lo que sólo nos resta incluir los apéndices indicados en el texto principal.

\section{Agradecimientos}

Queremos agradecer a Gary T. Horowitz, quien nos hizo notar por primera vez el porque de que Robert Wald se refería a los símbolos de Christoffel como un tensor. Agradecemos el apoyo del proyecto UNAM-DGAPA, PAPIIT IN 113115.

\section{Apéndice I: Sobre los co-vectores}\label{dual}

Para la presente discusión, supondremos que la noción de vector es familiar al lector, y haremos uso de ella para recordar en que consiste un co-vector\cite{Bretscher}.

Dado un espacio vectorial $V$ de dimensión finita, consideremos un espacio $V^*$ cuyos elementos son todas las posibles funciones lineales $v^*:V\rightarrow R$. Para los elementos de $V^*$, definimos el resultado de la suma y multiplicación por escalar como el elemento de $V^*$ que para todo elemento de $V$ cumple respectivamente con :
\begin{eqnarray}
[{v^*}^1+{v^*}^2](v)&=&{v^*}^1(v)+{v^*}^2(v),\nonumber \\  
\left[\alpha  {v^*}\right] (v) &=& \alpha \left[ {v^*}\right] (v).\label{linco}
\end{eqnarray}
Notamos que de estas dos definiciones se sigue que $V^*$ cumple con todas las propiedades de un espacio vectorial. Al espacio vectorial resultante se le conoce como el dual de $V$, y tiene el mismo número de dimensiones que este.

Una demostración formal respecto a la dimensionalidad de $V^*$ aporta poca intuición respecto a la naturaleza de este espacio, sin embargo una forma simple de convencerse de que $V^*$ tiene el mismo número de dimensiones que $V$ es como sigue. Consideremos un conjunto de $D$ vectores $e_i$ que constituyan una base del espacio $D$-dimensional $V$ y seleccionemos $D$ elementos $e^i$ de $V^*$ que cumplan con $e^i(e_j)={\delta^i}_j$. Usando las propiedades (\ref{linco}) es fácil ver que los elementos $e^i$ constituyen una base para $V^*$, pues sólo es necesario notar que toda función lineal $v^*:V\rightarrow R$ se puede escribir como una combinación lineal de todos los $e^i$ y ninguno de ellos se puede escribir como una combinación de los restantes. Dado que, tras verificar lo que se enuncia en la frase anterior, notamos que hemos encontrado una base para $V^*$ y esta tiene $D$ elementos, queda claro que la dimensionalidad de $V^*$ es la misma que la de $V$, y hemos ganado algo de intuición sobre el comportamiento de los elementos de $V^*$.

Otra forma de visualizar el espacio dual es como el mismo espacio vectorial provisto de un producto interno. Sabemos que el producto interno asocia de forma bilineal un número real a todo par de vectores, pues $v \cdot u\in R$. De esta forma podemos pensar en $v \cdot$ como un objeto listo para actuar sobre cualquier vector $u\in V$. El resultado de esta operación será un número real, y no sólo eso, sino que $v \cdot$ establecerá un mapeo lineal de $V$ a $R$, que será distinto para todo $v\in V$ que se elija para formarlo. En esta construcción es particularmente sencillo ver porque $V^*$ y $V$ son espacios tan parecidos.

La forma en que hemos descrito a $V^*$ es a través de sus constituyentes siendo objetos que actúan sobre los elementos de $V$ de la forma $v^*(u)=a\in R$, sin embargo, vemos que esta operación es simétrica, en el sentido de que a cada par constituido por un elemento de $V^*$ y uno de $V$ se asocia multilinealménte un real. Dada la simetría apenas descrita, vemos que es igualmente válido pensar en que un co-vector actúa sobre un vector o que un vector actúa sobre un co-vector. Es en este sentido que, así como $V^*$ es el espacio dual a $V$, es posible pensar en $V$ como isomorfo al espacio dual a $V^*$ y que sus elementos son mapeos lineales de $V^*$ a los reales.

\section{Apéndice II: Definición operacional de vector, conexiones sobre 1-formas y sobre campos tensoriales}\label{Apcon}

A lo largo del texto hemos supuesto que el lector está familiarizado con el álgebra vectorial y que conoce la definición de un vector como un elemento de un espacio que cumple con una serie de características. En el contexto de la relatividad general, es muy conveniente conocer a los vectores, o campos vectoriales, a través de una definición independiente de las coordenadas que se elijan y que refleje más cercanamente el espíritu con el que se escribe la relatividad general. Empecemos entonces por ver la definición operacional de un campo vectorial\cite{Baez:1995sj}.

Es común pensar en un vector como un objeto que dicta una dirección y que tiene una cierta magnitud. Es también familiar la noción de los operadores de derivada direccional sobre funciones, y en particular sabemos que esta dirección de diferenciación es especificada por un vector. Notemos ahora que, una derivada direccional al actuar sobre una función en algún punto del esapacio, resulta en un número, así que podemos pensar en una derivada direccional como un objeto que en cada punto del espacio base en que están definidas tanto ella como las funciones sobre las que actúa, establece un mapeo, $v_p$, del espacio de las funciones, $C^\infty$, a los reales. El hecho que nos es de interés, y que mencionamos sin demostrar, es que si se construye uno de estos mapeos $v_p$ por medio de especificar la acción que este tiene sobre todas y cada una de las funciones en $C^\infty$, entonces existe una única dirección y una única magnitud tal que la derivada direccional en esta dirección de toda función coincide con el resultado de aplicar este mapeo.

Vemos entonces que los mapeos $v_p$ están en correspondencia uno a uno con los vectores en el punto $p$ que especifican la dirección de derivación, y es por este motivo que se puede pensar en un vector tangente como un mapeo $v_p$.

Ahora, si tomamos un objeto $v$ que asocie un mapeo $v_p$ a cada punto $p$ del espacio, notamos que la acción $v(f)$ arroja un número $v_p(f)$ en cada punto $p$ del espacio, es decir, $v(f)$ es una función sobre el espacio.

Tomando en cuenta los comentarios anteriores, vemos que tiene sentido definir a un campo vectorial $v$ como un mapeo del espacio de las funciones $C^\infty$ en si mismo que cumple con
\begin{eqnarray}
v(f+g)&=&v(f)+v(g),\nonumber \\
v(\alpha\,f)&=&\alpha\,v(f),\nonumber \\
v(f\,\,g)&=&v(f)\,\, g+f\,\, v(g),\nonumber
\end{eqnarray}
con $f,\,\,g\in C^\infty$ y $\alpha$ un número real.

La definición anterior de $v$ tiene la ventaja de ser independiente de las coordenadas que se elijan, pues una función asocia un número a cada punto del espacio independientemente de las coordenadas que se usen para describirlo y lo mismo es cierto para la acción de $v$ sobre toda $f$.

Para hablar de la acción que tiene una conexión $D$ sobre tensores de diversos rangos, empecemos por mencionar que la acción de cualquier conexión sobre una función $f$ debe cumplir con $D_v(f)=v(f)$, y es por ello que al resultado $D_v(E)$ se le llama la derivada en la dirección $v$ del objeto $E$.

Hasta el momento hemos hablado de la acción de $D_v$ sobre funciones y campos vectoriales así que toca el turno de generalizar esta noción.

A partir de la definición de conexión sobre el haz tangente, es fácil determinar como es una conexión sobre el haz cotangente\cite{Baez:1995sj}, requiriendo que (\ref{leibniz}) se cumpla para productos más generales, y en particular que se cumpla
\begin{equation}
D_v[\omega( w_1)]=D_v(\omega)[w_1]+\omega[D_v(w_1)],\label{leibniz2}
\end{equation}
y aún más, dado que $\omega( w_1)$ es una función, y $v$ puede actuar directamente sobre ella, también se debe cumplir que
\begin{equation}
D_v[\omega( w_1)]=v[\omega( w_1)]. \label{dvf}
\end{equation}
Usando (\ref{leibniz2}) y (\ref{dvf}) la acción de $D_v$ sobre $\omega$ queda determinada.

Basados en la acción de la conexión sobre vectores, 1-formas y funciones, así como pidiendo que para campos tensoriales de rangos arbitrarios se cumpla que 
\begin{equation}
D_v[T_1\otimes T_2]=D_v[T_1]\otimes [T_2]+T_1\otimes D_v[T_2],
\end{equation}
se puede determinar la acción de la conexión sobre cualquier campo tensorial del rango que se desee\cite{Baez:1995sj}.

\bibliography{Bibliografía}

\end{document}